\documentclass[11pt,draftcls,onecolumn] {IEEEtran}

\usepackage[margin=1.9cm]{geometry}

\usepackage{amsmath}
\usepackage{amsfonts}
\usepackage{mathrsfs}
\usepackage{cite}
\usepackage{graphicx}
\usepackage{amssymb}
\usepackage{stfloats}
\usepackage{subeqnarray}%(1a)
\usepackage{cases}
\usepackage{indentfirst}

\begin{document}

\title{Capacity Analysis of Bidirectional AF Relay Selection with Imperfect Channel State Information}

\author{ Hongyu Cui, Rongqing Zhang, Lingyang Song, and Bingli Jiao
\thanks{H. Cui, R. Zhang, L. Song, and B. Jiao are with the
Electronics Engineering and Computer Science Department, Peking
University, Beijing, China, $100871$ (email:\{cuihongyu,
rongqing.zhang, lingyang.song, jiaobl\}@pku.edu.cn).} } \maketitle

\maketitle
\begin{abstract}
In this letter, we analyze the ergodic capacity of bidirectional
amplify-and-forward relay selection~(RS) with imperfect channel
state information~(CSI), i.e., outdated CSI and imperfect channel
estimation. Practically, the optimal RS scheme in maximizing the
ergodic capacity cannot be achieved, due to the imperfect CSI.
Therefore, two suboptimal RS schemes are discussed and analyzed, in
which the first RS scheme is based on the imperfect channel
coefficients, and the second RS scheme is based on the predicted
channel coefficients. The lower bound of the ergodic capacity with
imperfect CSI is derived in a closed-form, which matches tightly
with the simulation results. The results reveal that once CSI is
imperfect, the ergodic capacity of bidirectional RS degrades
greatly, whereas the RS scheme based on the predicted channel has
better performance, and it approaches infinitely to the optimal
performance, when the prediction length is sufficiently large.
\end{abstract}


\section{Introduction}
Recently, bidirectional relay network attracts a lot of interest,
because it has better spectral efficiency than conventional one-way
relay network when adopting the network coding
technique\cite{Popovshi2007}. In addition, the relay selection~(RS)
technique has been intensively researched in the bidirectional relay
network, due to its ability to achieve full diversity with a single
relay\cite{Jing2009,Song2011,Upadhyay2011,Kyu2009}. In
\cite{Jing2009,Song2011}, the symbol error rate~(SER) of
bidirectional RS was derived in a closed-form, which verifies that
RS can achieve full diversity. The ergodic capacity analysis of
bidirectional RS was obtained in \cite{Upadhyay2011,Kyu2009}.
Furthermore, imperfect channel state information~(CSI), i.e.,
outdated CSI and imperfect channel estimation, has great impact on
the performance of RS, which has been fully studied in one-way relay
network, such as the SER analysis \cite{Michalopoulos2010,Seyi2011},
and the ergodic capacity analysis \cite{Torabi2010,Chen2011}.
However, to the best of the authors' knowledge, all the previous
researches about bidirectional RS all assume the CSI is perfect, and
the impact of imperfect CSI, such as outdated CSI, on the
performance of bidirectional RS has not been investigated.

In this paper, we analyze the ergodic capacity of RS in
bidirectional amplify-and-forward~(AF) relay with imperfect CSI. The
system model and the imperfect CSI model are presented in Section
\uppercase\expandafter{\romannumeral2}. In Section
\uppercase\expandafter{\romannumeral3}, we discuss the optimal RS
scheme in maximizing the ergodic capacity. However, it cannot be
achieved practically, due to the imperfect CSI. Therefore, two
suboptimal RS schemes are analyzed, in which the first scheme is
based on the imperfect channel coefficients, and the second scheme
is based on the predicted channel coefficients. The tight lower
bound of the ergodic capacity for the suboptimal RS schemes is
derived in Section \uppercase\expandafter{\romannumeral4}. It is
noted that the analytical expression is provided under the
generalized network structure, i.e., the channel coefficients follow
independent but not necessarily identical complex-Gaussian fading,
and the correlation coefficients of outdated CSI are different for
different channels. In Section
\uppercase\expandafter{\romannumeral5}, Monte-Carlo simulations
verify that the analytical expression matches tightly with the
simulated results. The results reveal that the ergodic capacity of
bidirectional RS degrades greatly once CSI is imperfect, whereas the
RS scheme based on the predicted channel can compensate the
performance loss, and it approaches infinitely to the performance
achieved by the optimal RS scheme, when the prediction length is
sufficiently large.

\emph{Notation:}~$\left(\cdot\right)^*$, $\left(\cdot\right)^T$, and
$\left(\cdot\right)^H$ represent the conjugate, the transpose, and
the conjugate transpose, respectively. $\mathbb{E}$ is used for the
expectation and $\Pr$ represents the probability.

\section{System Model}

In this paper, we investigate a generalized bidirectional AF relay
network with two sources $S_j$, $j=1,2$, exchanging information
through $N$ relays $R_i$, $i=1,\ldots,N$, where each node is
equipped with a single half-duplex antenna. The transmit powers of
each source and each relay are denoted by $p_s$ and $p_r$,
respectively. The direct link between sources does not exist due to
the shadowing effect, and the channel coefficients between $S_j$ and
$R_i$ are reciprocal, denoted by $h_{ji}$. All the channel
coefficients are independent complex-Gaussian random variables~(RV)
with zero mean and variance of $\sigma_{h_{ji}}^2$, and these
coefficients are constant over the duration of one frame.

The whole procedure of bidirectional AF RS is divided into two parts
periodically: \emph{relay selection process} and \emph{data
transmission process}. During the relay selection process, e.g., the
$\tau$th frame, the central unit~(CU), i.e., $S_j$, selects the best
relay according to the predefined RS scheme, which will be discussed
in the next section. During the data transmission process, e.g., the
$\big(\tau+\tau'\big)$th frame, only the selected relay is used for
transmission, and other relays keep idle until the next relay
selection process comes. Due to the time-variation of channel, the
channel at the date transmission process
$h_{ji}\left(\tau+\tau'\right)$ is quite different from that at the
relay selection process $h_{ji}\left(\tau\right)$, and their
relationship is modeled as \cite{Michalopoulos2010}
\begin{align}\label{Eq:hf}
h_{ji}\left(\tau+\tau'\right)=\rho_{f_{ji}}h_{ji}\left(\tau\right)+\sqrt{1-{\rho_{f_{ji}}^2}}\varepsilon_{ji}
\end{align}
where $\varepsilon_{ji}$ is an independent identically distributed
RV with $h_{ji}\left(\tau\right)$; $\rho_{f_{ji}}=J_0\left(2\pi
f_{d_{ji}}T\tau'\right)$, where $J_0\left( \cdot \right)$ stands for
the zeroth order Bessel function\cite{Abramowitz}, $f_{d_{ji}}$ is
the Doppler spread, and $T\tau'$ is the time delay.

The imperfect channel estimation is also considered in this paper,
and the channel $h_{ji}$ and its estimate $\hat h_{ji}$ are related
by $h_{ji}=\hat h_{ji}+e_{ji}$\cite{Seyi2011}, in which $\hat
h_{ji}$ and the detection error $e_{ji}$ follow independent
zero-mean complex-Gaussian distributions with variances of
$\sigma_{\hat h_{ji}}^2$ and $\sigma_{e_{ji}}^2$, respectively, and
$\sigma_{e_{ji}}^2=0$ means no estimation error.

Considering the transmission via $R_i$, the data transmission
process of bidirectional AF relay is divided into two phases. During
the first phase, the sources simultaneously send their respective
information to $R_i$. The received signal at $R_{i}$ is $
r_{i}=\sqrt{p_s}h_{t,1i}s_1+\sqrt{p_s}h_{t,2i}s_2+n_{ri}$, where
$h_{t,ji}\buildrel \Delta \over=h_{ji}\left(\tau+\tau'\right)$,
$s_j$ denotes the modulated symbols transmitted by $S_j$ with the
average power normalized, and $n_{ri}$ is the additive white
Gaussian noise~(AWGN) at $R_i$, which is zero mean and variance of
$\sigma_n^2$. During the second phase, $R_i$ amplifies and forwards
the received signal back to the sources. The signal generated by
$R_i$ satisfies $t_i = \sqrt {p_r } \beta_i r_i$, where $\beta_i =
\left(p_{s}\hat \gamma_{t,1i}+ p_{s} \hat \gamma_{t,2i} + \sigma_n^2
\right)^{ -1/2} $ is the variable-gain
factor\cite{Michalopoulos2010}, and $\hat \gamma_{t,ji}= |\hat
h_{t,ji}|^2$. The received signal at $S_j$ via $R_i$ is $y_{ji} =
h_{t,ji}t_i + n_{sj}$, where $n_{sj}$ is the AWGN at $S_j$. After
reconstructing and canceling the self-interference, i.e., $\sqrt{p_s
p_r} \beta_i \hat h_{t,ji} \hat h_{t,ji}s_j$\cite{Song2011}, the
instantaneous received signal-plus-interference-to-noise
ratio~(SINR) at $S_j$ via $R_i$ is
\begin{equation}\label{Eq:gamma}
\gamma _{ji}  = \frac{{\psi _r \psi _s \hat \gamma _{t,ji} \hat
\gamma _{t,\bar ji} }}{{\left( {\tilde \psi _s  + \tilde \psi _r }
\right)\hat \gamma _{t,ji}  + \tilde \psi _s \hat \gamma _{t,\bar
ji}  + c}}
\end{equation}
where $\psi_s =p_s/ \sigma_n^2$, $\psi_r=p_r/\sigma_n^2$, $\tilde
\psi_s=\psi_s+\psi_s\psi_r\sigma_{e_{ji}}^2$, $\tilde
\psi_r=\psi_r+3\psi_s\psi_r\sigma_{e_{ji}}^2+\psi_s\psi_r\sigma_{e_{{\bar
j}i}}^2$, $c= 2\psi _r \psi _s \sigma _{e_{ji}}^4  +\psi _r \psi _s
\sigma _{e_{\bar ji}}^4 +\psi _r \sigma _{e_{ji}}^2 + 1$, and
$\{j,\overline j\}=\{1,2\}~\mbox{or}~\{2,1\}$.

\section{Relay Selection Schemes with Outdated CSI}

The ergodic capacity of the bidirectional RS is defined
as\cite{Upadhyay2011,Kyu2009}
\begin{align}\label{Eq:capacity}
\overline
C=\mathbb{E}_{\gamma_{1k},\gamma_{2k}}\left\{\frac{1}{2}\log_2\left(1+\gamma_{1k}\right)+\frac{1}{2}\log_2\left(1+\gamma_{2k}\right)\right\}
\end{align}
where $k$ in the index of the selected relay,  $\gamma_{jk}$ can be
obtained by \eqref{Eq:gamma}, and the pre-log factor $1/2$ means
that the transmission of one data block from one source to the other
occupies two phases.

To maximize the ergodic capacity, the optimal RS scheme is
\begin{align}\label{Eq:optimal}
k=\arg \max_i \min \left\{|\hat
h_{1i}\left(\tau+\tau'\right)|^2,|\hat
h_{2i}\left(\tau+\tau'\right)|^2\right\}
\end{align}
where the detailed explanation of the scheme is given in Appendix A.

However, $\hat h_{ji}\left(\tau+\tau'\right)$ is unknown at the
relay selection process, due to the time-variation of channel
\eqref{Eq:hf}. Accordingly, the optimal RS scheme \eqref{Eq:optimal}
cannot be achieved with outdated CSI, thus we discuss two suboptimal
RS schemes which can be implemented practically.

The first alternative RS scheme with outdated CSI is obtained by
substituting the channel $ \hat h_{ji}\left(\tau+\tau'\right)$ in
\eqref{Eq:optimal} with the outdated channel $\hat
h_{ji}\left(\tau\right)$ obtained at the relay selection
process\cite{Jing2009,Upadhyay2011}, i.e.,
\begin{equation}\label{Eq:BWC}
{k} = \arg \mathop {\max }\limits_i \min \left\{ \left| { \hat
h_{1i}\left(\tau\right) } \right|^2 ,\left| { \hat
h_{2i}\left(\tau\right) } \right|^2 \right\}.
\end{equation}

The second alternative RS scheme with outdated CSI is obtained by
substituting the channel $\hat h_{ji}\left(\tau+\tau'\right)$ in
\eqref{Eq:optimal} with its predicted value
$h_{p,ji}\left(\tau+\tau'\right)$, i.e.,
\begin{equation}\label{Eq:BWPC}
{k} = \arg \mathop {\max }\limits_i \min \left\{ \left| {
h_{p,1i}\left(\tau+\tau'\right) } \right|^2 ,\left| {
h_{p,2i}\left(\tau+\tau'\right) } \right|^2 \right\}.
\end{equation}

In this paper, the Wiener filter \cite{Haykin} is applied for
channel prediction, which is the linear optimal prediction in
minimizing the mean square error, thus we have $ h_{p,ji} \big(
{\tau+\tau'} \big) = {\bf{w}}_{opt,ji}^H {\bf{\tilde h}}_{ji}$,
where ${\bf{\tilde h}}_{ji}=\big[\hat h_{ji}\big(\tau\big), \hat
h_{ji}\big(\tau-\Delta\big),\cdots, \hat
h_{ji}\big(\tau-L_{ji}\Delta+\Delta\big)\big]^T$ contains the
current and previous $\left(L_{ji}-1\right)$ channel coefficients,
which are obtained from previous relay selection processes, $\Delta$
is the interval of adjacent relay selection processes in frames,
$L_{ji}$ is the prediction length, and
${\bf{w}}_{opt,ji}={{\bf{R}}_{ji}^{-1}} {\bf{r}}_{ji} $, in which
${\bf{R}}_{ji}=\mathbb{E}\left\{{\bf{\tilde h}}_{ji}{\bf{\tilde
h}}_{ji}^H\right\}$ and ${\bf{r}}_{ji}=\mathbb{E}\left\{{\bf{\tilde
h}}_{ji}{h_{t,ji}^{*}}\right\}$. According to \cite{Haykin}, $
h_{ji}\left(\tau+\tau'\right)=
h_{p,ji}\left(\tau+\tau'\right)+\sqrt{\sigma_{h_{ji}}^2-\sigma_{p_{ji}}^2}n_{ji}
$, where $h_{p,ji}\left(\tau+\tau'\right)$ is a complex-Gaussian RV
with zero mean and variance of
$\sigma_{p_{ji}}^2={\bf{r}}_{ji}^H{\bf{R}}_{ji}^{-1}{\bf{r}}_{ji}$,
$n_{ji}$ is an independent complex-Gaussian RV with zero mean and
unit variance. The normalized correlation coefficient between
$h_{p,ji}\left(\tau+\tau'\right)$ and
$h_{ji}\left(\tau+\tau'\right)$ satisfies
$\rho_{p_{ji}}=\sqrt{\sigma_{p_{ji}}^2/\sigma_{h_{ji}}^2}$.

For simplicity, the notation $\hat h_{s,ji}$ is used to represent $
\hat h_{ji}\left(\tau\right) $ and
$h_{p,ji}\left(\tau+\tau'\right)$. Specifically, for the RS scheme
\eqref{Eq:BWC}, $ \hat h_{s,ji} \buildrel \Delta \over =\hat
h_{ji}\left(\tau\right)$, and for the RS scheme \eqref{Eq:BWPC}, $
\hat h_{s,ji} \buildrel \Delta \over =
h_{p,ji}\left(\tau+\tau'\right)$, then $\hat \gamma_{s,ji}\buildrel
\Delta \over=| \hat h_{s,ji}|^2$. With the unified notation $\hat
h_{s,ji}$, the analysis of the RS schemes \eqref{Eq:BWC} and
\eqref{Eq:BWPC} can be expressed in a unified manner.

\section{Ergodic Capacity of Analysis}
To analyze the ergodic capacity of bidirectional AF RS in
\eqref{Eq:capacity}, the distribution functions of $\hat
\gamma_{t,jk}=\left| { \hat h_{j{k}} }
\left(\tau+\tau'\right)\right|^2$ in \eqref{Eq:gamma} need to be
obtained first.

\textbf{\emph{Lemma 1: }} According to the RS schemes and the
relationship between $\hat h_{t,ji}$ and $\hat h_{s,ji}$, the PDF of
$\hat \gamma_{t,j {k}}$ is expressed as
\begin{align}\label{Eq:lemma21}
f_{\hat \gamma _{t,jk} } \left( z \right) &= \sum\limits_{i = 1}^N
{\sum\limits_{t = 0}^{N - 1} {\sum\limits_{A_t } {\left( { - 1}
\right)^t \left( {1 + \sum\limits_{l \in A_t } {\frac{{\sigma
_{s,\bar ji}^2 }}{{\sigma _{s,l}^2 }}} } \right)} } } ^{ -
1}\Bigg[\frac{1}{{\sigma _{t,ji}^2 }}\exp \left( { -
\frac{z}{{\sigma _{t,ji}^2 }}} \right) + \frac{{\zeta _j }}{{\sigma
_{t,ji}^2 }}\exp \left( { -\frac{ \xi _j z}{\sigma _{t,ji}^2}
}\right)\Bigg]
\end{align}
where
\begin{align}\label{Eq:xi}
\xi _j  =\left( {\frac{1}{{\sigma _{s,i}^2 }} \hspace{-0.8mm}+
\hspace{-0.8mm}\sum\limits_{l \in A_t } {\frac{1}{{\sigma _{s,l}^2
}}} } \right)\left( {\frac{{\rho _{ji}^2 }}{{\sigma _{s,ji}^2
}}\hspace{-0.8mm} + \hspace{-0.8mm}\frac{{1\hspace{-0.8mm} -
\hspace{-0.8mm}\rho _{ji}^2 }}{{\sigma _{s,i}^2 }}
\hspace{-0.8mm}+\hspace{-0.8mm} \sum\limits_{l \in A_t }
{\frac{{1\hspace{-0.8mm} -\hspace{-0.8mm} \rho _{ji}^2 }}{{\sigma
_{s,l}^2 }}} } \right)^{ - 1},
\end{align}
\begin{align}\label{Eq:zeta}
\zeta _j  = \left( {\frac{{\sigma _{s,\bar ji}^2 }}{{\sigma
_{s,ji}^2 }}\sum\limits_{l \in A_t } {\frac{1}{{\sigma _{s,l}^2 }}}
} \right)\left( {\frac{{\rho _{ji}^2 }}{{\sigma _{s,ji}^2
}}\hspace{-0.8mm} + \hspace{-0.8mm}\frac{{1
\hspace{-0.8mm}-\hspace{-0.8mm} \rho _{ji}^2 }}{{\sigma _{s,i}^2 }}
+ \sum\limits_{l \in A_t } {\frac{{1 \hspace{-0.8mm}-\hspace{-0.8mm}
\rho _{ji}^2 }}{{\sigma _{s,l}^2 }}} } \right)^{ - 1}.
\end{align}

In addition, $\sum_{A_t }$ is the abbreviation of
$\sum_{\scriptstyle A_t \subseteq \left\{ {1, \ldots ,N}
\right\}\backslash i \hfill \atop
 \scriptstyle \left| {A_t } \right| = t \hfill}$, and $\left| {A_t } \right|$
represents the cardinality of set $A_t$. Moreover, $\sigma_{s,ji}^2$
and $\sigma_{t,ji}^2$ are the variances of $\hat h_{s,ji}$ and $\hat
h_{t,ji}$, respectively. Specifically, for the RS scheme
\eqref{Eq:BWC}, $\sigma_{s,ji}^2=\sigma_{\hat h_{ji}}^2$,
$\sigma_{t,ji}^2=\sigma_{\hat h_{ji}}^2$,
$\rho_{ji}=\rho_e\rho_{f_{ji}}$, and $\rho_e=\sigma_{\hat
h_{ji}}^2/\sigma_{h_{ji}}^2$\cite{Seyi2011}; for the RS scheme
\eqref{Eq:BWPC}, $\sigma_{s,ji}^2=\sigma_{p_{ji}}^2$,
$\sigma_{t,ji}^2= \sigma_{\hat h_{ji}}^2$, and
$\rho_{ji}=\rho_e\rho_{p_{ji}}$. Also,
$\sigma_{s,i}^2=\sigma_{s,1i}^2\sigma_{s,2i}^2/\big(\sigma_{s,1i}^2+\sigma_{s,2i}^2\big)$.

\emph{\textbf{Proof: }} The derivation is given in Appendix B.
$\hfill \blacksquare$

\textbf{\emph{Lemma 2: }}
\begin{align}\label{Eq:eq3}
& \Theta \left( {a,m,n} \right) \buildrel \Delta \over =
\int\limits_0^\infty {\int\limits_0^y {\ln \left( {y + a} \right)e^{
- m\left( {y - x} \right)} e^{ - nx} dxdy} }= \left\{
{\begin{array}{*{20}l}
   {\frac{{\varphi \left( {a,n} \right) - \varphi \left( {a,m} \right)}}{{m - n}},m \ne n;}  \\
   {\frac{{1 - ame^{am} E_1 \left( {am} \right) + m\varphi \left( {a,m}
\right)}}{{m^2 }}
,m = n.}\\
\end{array}} \right.
\end{align}
where $\varphi \left( {a,b} \right) \buildrel \Delta \over =
\int\limits_0^\infty  {\ln \left( {x + a} \right)e^{ - bx} dx}  =
\frac{{\ln a + e^{ab} E_1 \left( {ab}
\right)}}{b}$\cite[4.337]{Gradshteyn94}, and $E_1\left(x\right)$ is
the exponential integral function \cite{Abramowitz}.

\emph{\textbf{Proof: }} Lemma~2 can be achieved by applying the
integration by part, then discussing under the situations that $m=n$
and $m \ne n$. $\hfill \blacksquare$

\textbf{\emph{Proposition~1 :}} Applying the Lemma~1 and Lemma~2,
the ergodic capacity of bidirectional AF RS is
\begin{align}\label{Eq:C}
\overline C=\frac{T_1+T_2-T_3-T_4}{2\ln2}
\end{align}
where
\begin{align}\label{Eq:T1}
T_j&=\hspace{-0.8mm} \sum\limits_{i = 1}^N {\sum\limits_{t = 0}^{N -
1} {\sum\limits_{A_t } {\left( {1\hspace{-0.8mm} +
\hspace{-0.8mm}\sum\limits_{l \in A_t } {\frac{{\sigma _{s,\bar
ji}^2 }}{{\sigma _{s,l}^2 }}} } \right)^{ - 1} \frac{{\left( { - 1}
\right)^t }}{{\sigma _{t,ji}^2 }}} } } \Bigg[\frac{1}{{\psi _r
}}\varphi \left( {m,\frac{1}{{\psi _r \sigma
_{t,ji}^2 }}} \right) \notag\\
&+\hspace{-0.8mm} \frac{{\zeta _j }}{{\psi _r }}\varphi \left(
{m,\frac{{\xi _j }}{{\psi _r \sigma _{t,ji}^2}}} \right)
\hspace{-0.8mm}+\hspace{-0.8mm} \frac{1}{{\psi _s }}\varphi \left(
{n ,\frac{1}{{\psi _s \sigma _{t,ji}^2 }}} \right)
\hspace{-0.8mm}+\hspace{-0.8mm} \frac{{\zeta _j }}{{\psi _s
}}\varphi \left( {n,\frac{{\xi _j }}{{\psi _s \sigma _{t,ji}^2}}}
  \right)\Bigg],
\end{align}
\begin{align}\label{Eq:T3}
T_{j + 2} & = \frac{1}{{\left( {\tilde \psi _s  + \tilde \psi _r } \right)\tilde \psi _s }}\sum\limits_{i = 1}^N {\sum\limits_{t = 0}^{N - 1} {\sum\limits_{A_t } {\sum\limits_{i' = 1}^N {\sum\limits_{t' = 0}^{N - 1} {\sum\limits_{A_{t'} } {\frac{{\left( { - 1} \right)^t }}{{\sigma _{t,ji}^2 }}} } } } } } \frac{{\left( { - 1} \right)^{t'} }}{{\sigma _{t,\bar ji'}^2 }} \left( {1 + \sum\limits_{l \in A_t } {\frac{{\sigma _{s,\bar ji}^2 }}{{\sigma _{s,l}^2 }}} } \right)^{ - 1} \left( {1 + \sum\limits_{l' \in A_{t'} } {\frac{{\sigma _{s,ji'}^2 }}{{\sigma _{s,l'}^2 }}} } \right)^{ - 1}  \notag\\
&\times \hspace{-0.8mm}\Bigg[\Theta \left( {mn,\frac{{1/\sigma _{t,ji}^2 }}{{\tilde \psi _s  \hspace{-0.8mm}+\hspace{-0.8mm} \tilde \psi _r }},\frac{{1/\sigma _{t,\bar ji'}^2 }}{{\tilde \psi _s }}} \right) \hspace{-0.8mm}+ \hspace{-0.8mm}\zeta '_{\bar j} \Theta \left( {mn,\frac{{1/\sigma _{t,ji}^2 }}{{\tilde \psi _s  \hspace{-0.8mm}+\hspace{-0.8mm} \tilde \psi _r }},\frac{{\xi '_{\bar j} /\sigma _{t,\bar ji'}^2 }}{{\tilde \psi _s }}} \right) \notag\\
&+ \hspace{-0.8mm}\zeta _j \Theta \left( {mn,\frac{{\xi _j /\sigma
_{t,ji}^2 }}{{\tilde \psi _s  \hspace{-0.8mm}+\hspace{-0.8mm} \tilde
\psi _r }},\frac{{1/\sigma _{t,\bar ji'}^2 }}{{\tilde \psi _s }}}
\right) \hspace{-0.8mm}+ \hspace{-0.8mm}\zeta _j \zeta '_{\bar j}
\Theta \left( {mn,\frac{{\xi _j /\sigma _{t,ji}^2 }}{{\tilde \psi _s
\hspace{-0.8mm}+\hspace{-0.8mm} \tilde \psi _r }},\frac{{\xi '_{\bar
j} /\sigma _{t,\bar ji'}^2 }}{{\tilde \psi _s }}} \right)\Bigg].
\end{align}

In addition, $m=\tilde \psi_s/\psi_s$, and $n=\big(\tilde \psi_s +
\tilde \psi_r\big)/\psi_r$. $\xi'_j$ and $\zeta'_j$ can be obtained
by $\xi_j$ and $\zeta_j$, respectively, by substituting $i, l, A_t$
in \eqref{Eq:xi} and \eqref{Eq:zeta} with $i', l', A_{t'}$,
respectively, and
 $\{j,\overline j\}=\{1,2\}~\mbox{or}~\{2,1\}$.

 It is noted that for the symmetric network structure, i.e.,
$\sigma_{h_{ji}}^2=1$, $\rho_{ji}=\rho$, $i=1,\ldots,N$, $j=1,2$, we
have $\sum_{A_t}=\binom{N-1}{t}$, and $\sum_{A_t'}=\binom{N-1}{t'}$,
thus the expression of capacity in Proposition 1 can be further
simplified.

\emph{\textbf{Proof: }} The derivation is given in Appendix C.
$\hfill \blacksquare$

\section{Simulation Results and Discussion}
In this section, Monte-Carlo simulations are provided to validate
the preceding analysis and to highlight the performance of
bidirectional AF RS with outdated CSI. Without loss of generality,
 the sources and the relays are assumed to have the same transmit powers, i.e.,
$p_s=p_r=P_0$. The network structure is assumed to be symmetric,
i.e., $\sigma_{h_{ji}}^2=1$ and $f_{d_{ji}}=f_d$, $j=1,2$,
$i=1,\ldots,N$.

Fig.~1 studies the impact of outdated CSI on the ergodic capacity
when $N=4$ and adopting the RS scheme \eqref{Eq:BWC}. The x-axis is
$\mbox{SNR}=P_0/\sigma_n^2$, and the imperfect estimation is
considered, i.e., the variance of detection error
$\sigma_e^2=\sigma_n^2/P_0$\cite{Ikki2012}. Different lines are
provided under different $f_dT$, where larger $f_dT$ means CSI is
outdated more severely than smaller $f_dT$, and $f_dT=0$ means CSI
is not outdated. The figure verifies that the expression of
Proposition~1 is the tight lower bound of the simulated results. We
also observe that the ergodic capacity degrades once CSI is
outdated, e.g., the performance loss between $f_dT=0.3$ and $f_dT=0$
is about $4$ dB in high SNR, and more severely outdated CSI results
in greater performance loss, although outdated CSI has no impact on
the multiplexing gain.

Figs.~2-4 study the capacity of bidirectional AF RS with outdated
CSI, when adopting the RS scheme \eqref{Eq:BWPC} and the variance of
detection error $\sigma_e^2=0$. We assume the time interval between
relay selection process and its subsequent data transmission process
satisfies $\tau'=1$, and the time interval between adjacent relay
selection processes satisfies $\Delta=2$. The prediction lengths of
different channels are assumed to be the same, i.e., $L_{ji}=L$,
$j=1,2$, $i=1,\ldots,N$.

Fig.~2 plots the simulated and the analytical ergodic capacity of
outdated CSI versus $\mbox{SNR}=P_0/\sigma_n^2$, when the
coefficient of outdated CSI is fixed, i.e., $f_dT=0.3$. The
simulated results verify the lower bound in Proposition~1 is tight
when the channel prediction is adopted. For the symmetric network,
the line of $L=1$ represents the RS scheme \eqref{Eq:BWC}, because
$\rho_{p}=\rho_{f}$ when $L=1$. Also, the line of $L=\infty$ means
the CSI is perfect, because $\rho_p=1$ when $L=\infty$. As this
figure reveals, the RS scheme based on channel prediction
\eqref{Eq:BWPC} outperforms the scheme without channel prediction
\eqref{Eq:BWC}, and the performance gain gets larger when increasing
the prediction length. The qualitative explanation of the phenomenon
is that increasing the prediction length results in growing the
correlation coefficient $\rho_{p}$, thus the prediction becomes more
accurate and the performance gets improved.

Fig.~3 investigates the impact of $f_dT$ on capacity when
$\mbox{SNR}=10$, $15$, and~$20$~dB. As the figure reveals, the
curves of capacity under different prediction lengths $L$ have
almost the same performance in small $f_dT$, and all the capacity
degrades as $f_dT$ increases. However, larger $L$ has better
robustness of capacity, e.g., for $L=1$ and SNR~=~$15$ dB, capacity
degrades to $3$ bps/Hz when $f_dT=0.18$, whereas for $L=2$, it is
$f_dT=0.27$ when capacity degrades to the same level. Therefore,
larger $L$ improves the robustness of capacity.

Fig.~4 investigates the impact of prediction length $L$ on the
capacity. The y-axis is the normalized difference of capacity, i.e.,
the difference between the capacity obtained by the RS scheme
\eqref{Eq:BWPC} and the capacity obtained by the optimal RS scheme
\eqref{Eq:optimal}, normalized by the latter. Smaller normalized
difference means the capacity of the RS scheme \eqref{Eq:BWPC} with
outdated CSI has closer performance with the optimal performance.
Different lines of Fig.~4 are plotted under different $f_dT$ and
SNR, and all the normalized differences decrease monotonously to
zero as $L$ increases, which reveals that the RS scheme
\eqref{Eq:BWPC} can infinitely approach to the optimal performance,
by enlarging the prediction length. Furthermore, the results also
reveal that the lines with larger $f_dT$ needs larger $L$ to satisfy
the same need of the normalized difference.

\section{Conclusions}
The impact of imperfect CSI on the ergodic capacity of bidirectional
AF RS has been investigated in this paper. The tight lower bound of
ergodic capacity is derived in a closed form and verified by
simulations. The results reveal that imperfect CSI will jeopardize
the capacity of bidirectional RS network, whereas the RS based on
the channel prediction can compensate the performance loss, and it
can infinitely approach to the optimal performance, by enlarging the
prediction length.

\section*{Appendix A: Explanation of the Optimal Scheme}

To maximize the ergodic capacity of \eqref{Eq:capacity} is
equivalent to maximize $\gamma_{1i}\gamma_{2i}$, according to the
approximation $\log_2\big(1+\gamma_{ji}\big) \approx
\log_2\gamma_{ji}$, in which $\gamma_{ji}$ is provided in
\eqref{Eq:gamma}. Using the approximation $xy/\big(x+y+c\big)\approx
\min\big(x,y\big)$, $\gamma_{1i}\gamma_{2i}$ is bounded by $\min
\big(|\hat h_{t,1i}|^2,|\hat h_{t,2i}|^2\big)^2$. Therefore,
\eqref{Eq:capacity} is optimal in maximizing the capacity.

\section*{Appendix B: Proof of Lemma~1 }
Similar to \cite{Michalopoulos2010}, the PDF of $\hat \gamma_{t,1k}$
can be expressed as
\begin{align}\label{Eq:App1}
f_{\hat\gamma _{t,1k} } \left( z \right)& \buildrel (a) \over =
\frac{d}{{dz}}\sum\limits_{i = 1}^N {\Pr \left\{ {\hat \gamma
_{t,1i} <
z \cap k = i} \right\}} \notag\\
&\buildrel (b) \over = \sum\limits_{i = 1}^N {\int\limits_0^\infty  {f_{\hat\gamma _{t,1i} |\hat \gamma _{s,1i} } \left( {z|y} \right)f_{\hat \gamma _{s,1i} } \left( y \right)} } \Pr \left\{ {\hat \gamma _{s,1i}  \le \hat \gamma _{s,2i} |\hat \gamma _{s,1i}  = y} \right\} \underbrace {\Pr \left\{ {k = i|\hat \gamma _{s,1i}  = y,\hat \gamma _{s,1i}  \le \hat \gamma _{s,2i} } \right\}}_{I_1 }dy \notag\\
&+\hspace{-1mm}\sum\limits_{i = 1}^N
{\hspace{-1mm}\int\limits_0^\infty\hspace{-1.4mm} {f_{\hat\gamma
_{t,1i} |\hat \gamma _{s,1i} } \left( {z|y} \right)f_{\hat \gamma
_{s,1i} } \left( y \right)} } \Pr \left\{ {\hat \gamma
_{s,1i}\hspace{-1mm}
> \hspace{-1mm}\hat \gamma _{s,2i} |\hat \gamma _{s,1i} = y}
\right\}\underbrace {\Pr \left\{ {k\hspace{-1mm} =\hspace{-1mm}i
|\hat \gamma _{s,1i}\hspace{-1mm} =\hspace{-1mm} y,\hat \gamma
_{s,1i} \hspace{-1mm}
>\hspace{-1mm} \hat \gamma _{s,2i} } \right\}}_{I_2 }dy
\end{align}
where (a) is satisfied by the total probability theorem, which
divides the union event $\hat \gamma_{t,1k}<z$ into $N$ disjoint
events, i.e., $R_i$ is the best relay and $\hat \gamma_{t,1i}<z$,
$i=1,\ldots,N$; (b) is fulfilled by the division of two disjoint
events, i.e., $\hat \gamma_{s,1i}> \hat \gamma _{s,2i}$ and $ \hat
\gamma _{s,1i} \le \hat \gamma _{s,2i}$. Furthermore, $I_1$ and
$I_2$ can be obtained by the definition of the RS schemes
\eqref{Eq:BWC} and \eqref{Eq:BWPC}, the order statistics, and
\cite[eq.~26]{Hai2011}. Substituting the exponential distributions
of $ \hat \gamma_{s,1i}$ and $\hat  \gamma_{s,2i}$ into
\eqref{Eq:App1}, Lemma 1 is verified by applying the conditional PDF
of $\hat \gamma_{t,1k}$ and $\hat \gamma_{s,1k}$ and
\cite[eq.~6.614.3]{Gradshteyn94}. Moreover, the PDF of $\hat
\gamma_{t,2k}$ can be obtained similarly.

\section*{Appendix C: Proof of Proposition~1}
The ergodic capacity of bidirectional RS is tightly bounded by
\begin{align}\label{Eq:appr}
&\overline C \buildrel (a) \over\ge \mathbb{E}_{\hat\gamma _{t,1k} ,\hat\gamma _{t,2k} } \left\{ {\frac{1}{2}\log _2 \left[ {\frac{{\left( {\psi_r \hat\gamma _{t,1k}\hspace{-0.8mm}  +\hspace{-0.8mm} m} \right)\left( {\psi_s \hat\gamma _{t,2k}\hspace{-0.8mm}  +n} \right)}}{{\left(\tilde \psi_s\hspace{-0.8mm}+\hspace{-0.8mm}\tilde \psi_r\right) \hat \gamma _{t,1k} \hspace{-0.8mm} + \hspace{-0.8mm}\tilde \psi_s \hat \gamma _{t,2k}\hspace{-0.8mm}  +mn}}} \right]} \right\} \notag\\
&+ \mathbb{E}_{\hat \gamma _{t,1k} ,\hat \gamma _{t,2k} } \left\{
{\frac{1}{2}\log _2 \left[ {\frac{{\left( {\psi_s \hat\gamma _{t,1k}
\hspace{-0.8mm} +n} \right)\left( {\psi_r \hat \gamma
_{t,2k}\hspace{-0.8mm}  + m} \right)}}{{\tilde\psi_s \hat\gamma
_{t,1k} \hspace{-0.8mm} +\hspace{-0.8mm}
\left(\tilde\psi_s\hspace{-0.8mm}+\hspace{-0.8mm}\tilde\psi_r\right)
\hat\gamma _{t,2k}\hspace{-0.8mm}  +m n}}} \right]} \right\}
\end{align}
where $m=\tilde \psi_s/\psi_s$, and $n=\big(\tilde \psi_s + \tilde
\psi_r\big)/\psi_r$. Comparing with \eqref{Eq:gamma}, $(a)$ is
achieved by adding the constant $mn-c$ in the dominator of
\eqref{Eq:appr}, which has little effect on the performance, like
the SER analysis \cite{Song2011}.

After some manipulation of the logarithmic and expectation
operations, the capacity of \eqref{Eq:appr} is rewritten as $
\overline C=\frac{T_1+T_2-T_3-T_4}{2\ln 2}$, where
$T_1={\mathbb{E}_{\hat \gamma _{t,1k} } \big\{ {\ln \left( {\psi_r
\hat \gamma _{t,1k} +m} \right)\hspace{-0.8mm} + \hspace{-0.8mm}\ln
\left( {\psi_s \hat\gamma _{t,1k}\hspace{-0.8mm} +n} \right)}
\big\}}$, and $T_2$ can be obtained by substituting $\hat
\gamma_{t,1k}$ in $T_1$ with $\hat \gamma_{t,2k}$. According to the
fact that $f_Y\left(z\right)=f_X\left(z/m\right)/m$ when
$Y=mX\left(m>0\right)$, the PDFs of $\psi_s\hat\gamma_{t,jk}$ and
$\psi_r\hat\gamma_{t,jk}$ can be obtained by the PDF of
$\hat\gamma_{t,jk}$, $j=1,2$, then \eqref{Eq:T1} can be achieved by
Lemmas~1 and 2. Moreover, $T_3={\mathbb{E}_{\hat\gamma _{t,1k}
,\hat\gamma _{t,2k} } \bigg\{ {\ln \Big[
{\left(\tilde\psi_s\hspace{-0.8mm}+\hspace{-0.8mm}\tilde\psi_r\right)
\hat\gamma _{t,1k}\hspace{-0.8mm}  +\hspace{-0.8mm} \tilde\psi_s
\hat\gamma _{t,2k}\hspace{-0.8mm}  +mn} \Big]} \bigg\}}$, and $T_4$
can be obtained by permuting $\hat \gamma_{t,1k}$ with $\hat
\gamma_{t,2k}$ in $T_3$. $\hat \gamma_{t,1k}$ and $\hat
\gamma_{t,2k}$ are independent RVs, thus the PDF of
$\left(\tilde\psi_s+\tilde\psi_r\right)\hat\gamma_{t,1k}+\tilde\psi_s\hat\gamma_{t,2k}$
can be achieved by the convolution of
$\left(\tilde\psi_s+\tilde\psi_r\right)\hat\gamma_{t,1k}$'s and
$\tilde\psi_s\hat\gamma_{t,2k}$'s PDFs, then $T_3$ in \eqref{Eq:T3}
can also be achieved by Lemmas~1 and 2. The expression of $T_4$ can
be obtained similarly.

\newpage

\begin{figure}
\centering
\includegraphics[height=3.5in,width=3.8in]{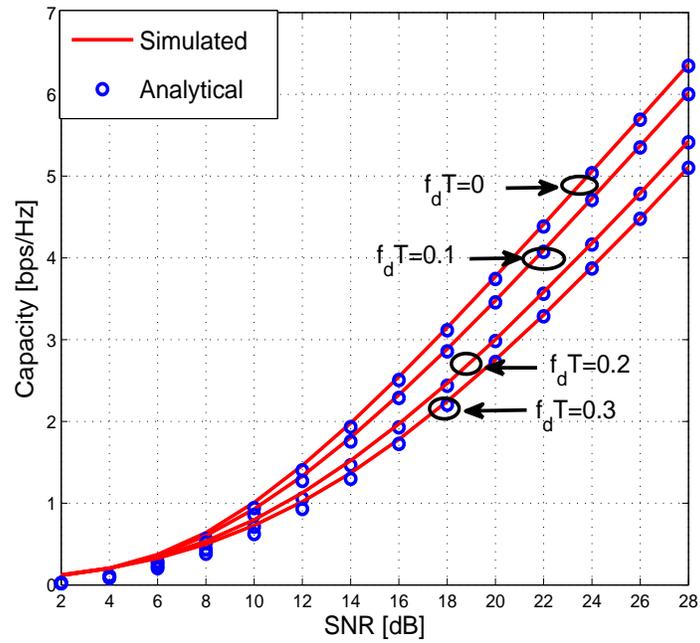}
\caption{Capacity under outdated CSI when applying the scheme
\eqref{Eq:BWC}.}
\end{figure}

\begin{figure}
\centering
\includegraphics[height=3.5in,width=3.8in]{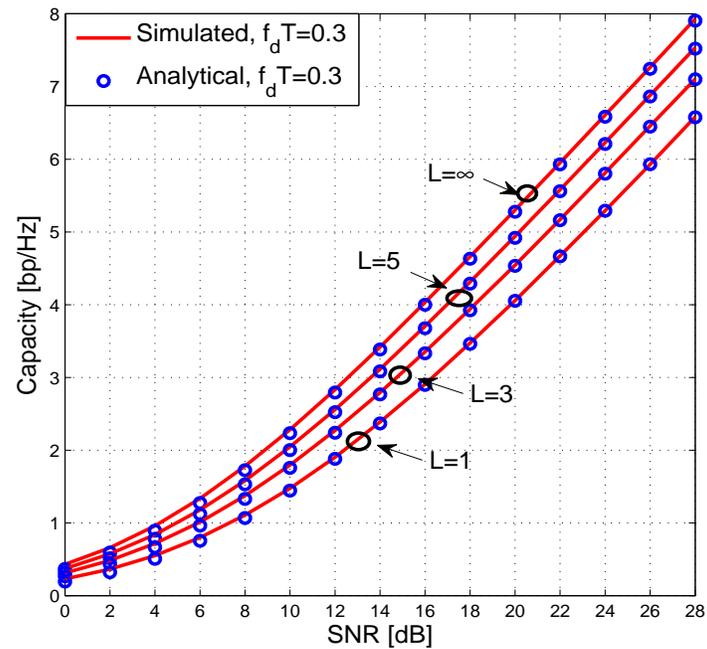}
\caption{Capacity when applying the RS scheme \eqref{Eq:BWPC}.}
\end{figure}

\begin{figure}
\centering
\includegraphics[height=3.5in,width=3.8in]{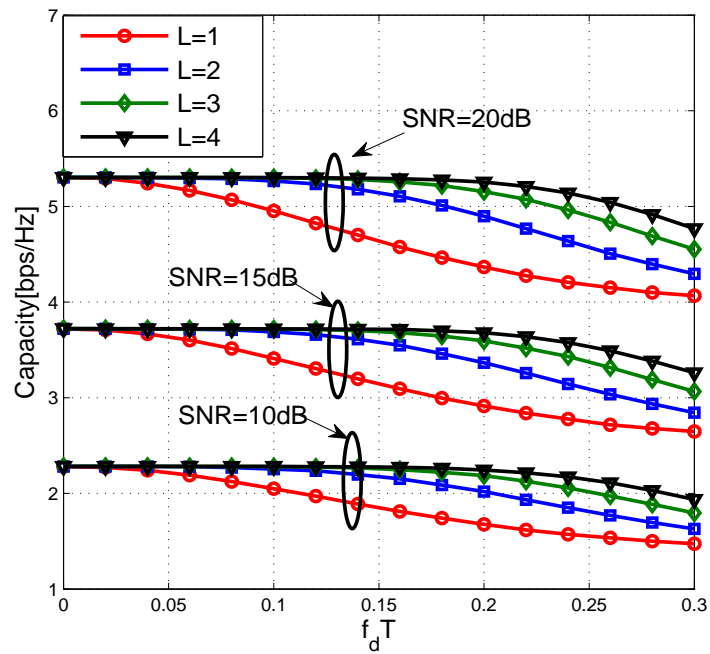}
\caption{The impact of $f_dT$ on the capacity.}
\end{figure}

\begin{figure}
\centering
\includegraphics[height=3.5in,width=3.8in]{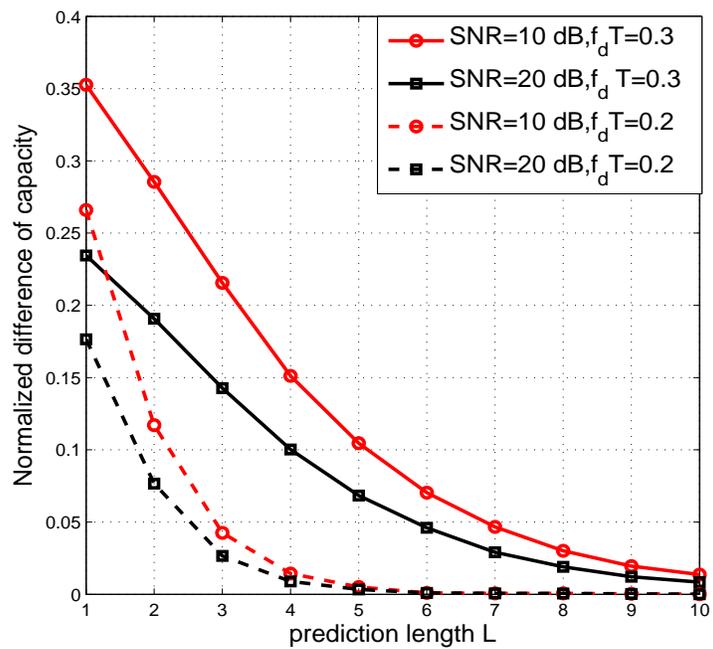}
\caption{The normalized difference of capacity versus prediction
length $L$.}
\end{figure}

\end{document}